\documentclass[a4paper,12pt]{article}

\usepackage{graphicx}
\usepackage{amsmath,amssymb}
\usepackage{a4wide}
\usepackage{cite}

\renewcommand{\baselinestretch}{1.5}

\begin{document}

{ \renewcommand{\baselinestretch}{1.4}\small\normalsize

\begin{center}

   {\large \bf Stress Development and Impurity Segregation \\
               during Oxidation of the Si(100) Surface}

   \vspace{0.5cm}
    Daniel J. Cole,$^{1\ast}$
    Mike C. Payne,$^1$ and Lucio Colombi Ciacchi$^{2,3}$

   \vspace{0.25cm}
   {\it
       $^1$Theory of Condensed Matter Group, Cavendish Laboratory, \\
           University of Cambridge, J J Thomson Avenue, Cambridge CB3 0HE, UK\\
       $^2$Fraunhofer Institut f\"ur Werkstoffmechanik \\
           W\"ohlerstrasse 11, 79108 Freiburg, Germany \\
       $^3$Institut f\"ur Zuverl\"assigkeit von Bauteilen und Systemen, \\
           University of Karlsruhe, Kaiserstr. 12, 76131 Karlsruhe, Germany. }

\end{center}

\date{}

\begin{abstract}
We have studied the segregation of P and B impurities during 
oxidation of the Si(100) surface by means of combined static and
dynamical first-principles simulations based on density functional
theory.
In the bare surface, dopants segregate to chemically stable surface
sites or to locally compressed subsurface sites.
Surface oxidation is accompanied by development of tensile surface
stress up to 2.9~N/m at a coverage of 1.5 monolayers of oxygen and by
formation of oxidised Si species with charges increasing approximately
linearly with the number of neighbouring oxygen atoms.
Substitutional P and B defects are energetically unstable within the
native oxide layer, and are preferentially located at or beneath the
Si/SiO$_x$ interface.
Consistently, first-principles molecular dynamics simulations of
native oxide formation on doped surfaces reveal that dopants avoid the
formation of P--O and B--O bonds, suggesting a surface oxidation
mechanism whereby impurities remain trapped at the Si/SiO$_x$
interface.
This seems to preclude a direct influence of impurities on the surface
electrostatics and, hence, on the interactions with an external
environment.
\end{abstract}


\vfill \noindent
$^{\ast}$To whom correspondence should be addressed;\\
e-mail: djc56@cam.ac.uk; phone: +44 (0)1223 337049; fax: +44 (0)1223 337356

}  % \renewcommand{\baselinestretch}{1.4}\small\normalsize

\section{Introduction}

Interactions between Si-based microelectromechanical systems (MEMS)
and their external environment are particularly important when their
surfaces are in contact with an atmospheric or physiological
milieu~\cite{bioactivity,biomems,physiol,immobilisation}.
These interactions, largely of a non-covalent type, directly involve 
the natively oxidised Si surface and are thus strongly dependent on 
the distribution of charged species within the oxide layer.
Such charge distributions may be affected by P and B dopants present 
in the Si bulk, if they tend to segregate into the oxide layer.
Experimental and theoretical investigations~\cite{STMphosphine,Ramamoo} 
have established that P and B dopants segregate to the bare Si(100) surface in 
order to saturate dangling bonds and relieve strain in the Si lattice 
caused by size mismatch between the Si and dopant atoms.
This combination of chemical and mechanical effects is expected to have a 
similar effect on impurity segregation into the thin native oxide surface layer that 
spontaneously forms on Si in the presence of oxygen. 
However, the detailed behaviour of such dopants during and after the 
oxidation process remains unclear and is currently the subject of a number
of theoretical and experimental studies.

To address the issue of P segregation at the Si/SiO$_2$ interface, 
secondary ion mass spectroscopy (SIMS) has been widely used since it is
capable of measuring compositions varying over several orders of
magnitude as a function of depth.
SIMS experiments reveal a decrease in P dopants of up to 50$\%$
on removal of the oxide layer, although the limited resolution in this
case makes it impossible to determine whether these dopants were
present in the oxide layer or at the interface~\cite{Griffin}.
On the other hand, inductively coupled plasma mass spectrometry
measured just 2$\%$ of the initial dose of P atoms in the oxide layer
after oxidation at 850~$^{\circ}$C, with the majority shown by precise
SIMS profiles piled up on the Si side of the interface~\cite{icpms}.
These results are consistent with ab-initio simulations of such
interfaces~\cite{dabrow1,physica2}, in which it is found that P atoms
avoid forming P--O bonds and are stable at threefold-coordinated
defect sites, introduced during the oxidation process.
At dopant concentrations above 0.01~nm$^{-3}$, P atom pairs dominate
with the interface accommodating the stress induced by bond
deformation around the resulting defect.

In contrast to the Si/SiO$_2$ dielectric interface, there have been
relatively few studies of impurity segregation at the interface between
Si and its ultrathin native oxide.
Such an interface contains a high concentration of defects 
(approximately 0.1~nm$^{-2}$ \cite{dabrow1}), in the form
of Si dangling bonds.
SIMS experiments on a Si(100) native oxide report P segregation into 
the oxide layer at room temperature, which 
increases with annealing \cite{Ying4}.
However, these measurements are taken at very high doping
concentration (0.5~nm$^{-3}$) and possible pile-up at the interface
cannot be observed since the oxide layer is thin compared to the SIMS
resolution.
Auger electron spectroscopy data of a similar interface showed no
spectral line corresponding to oxidised P \cite{dabrow1}.

There has been similar uncertainty among studies addressing the 
segregation behaviour of B impurities.
A theoretical study of the Si/SiO$_2$ interface has proposed that B
segregates to threefold-coordinated O vacancies in the oxide layer
\cite{furu-theor}.
This is in response to experimental observations, reporting B diffusion
through thin oxide layers during annealing of doped polysilicon
\cite{fair-diffuse}, and SIMS profiles, revealing a sixfold excess of
dopant atoms on the oxide side of the interface compared with the Si
side \cite{sims-1984}.
On the other hand, recent reports suggest that surface peaks in B
doped samples are artefacts of the SIMS analysis and are not confirmed
by other techniques, known to be accurate at high doping concentration
\cite{SIMS-no-good}. 
An energy-filtered transmission electron microscopy B profile of a
clean Si(100) surface, doped at a concentration of 50~nm$^{-2}$, shows
a peak at a depth of 60~\AA{} \cite{eftem-B}.
The sample was shown to have a native oxide depth of approximately
15~\AA{}, placing the B peak in the strained Si layer below the
interface.
This result is supported by high-resolution Rutherford backscattering
spectroscopy, revealing that B is concentrated beneath the native
oxide layer, formed after B implantation \cite{hrbs-B}.

In this work, we aim to gain a basic understanding of the behaviour of 
P and B impurities during the formation of a native oxide layer on the
Si(100) surface by performing combined static and dynamical first-principles
simulations based on Density Functional Theory (DFT).
Both the changes in Si coordination and the evolution of surface stress that 
occur during the formation of the native oxide layer on the bare p($2\!\times\!2$)
reconstructed Si(100) surface will be investigated.
We will then perform total energy calculations of dopant substitution into 
various sites of bare and natively oxidised surface models.
These static calculations are able to discriminate the thermodynamically stable 
dopant positions, but are limited by the number of configurations that may be 
investigated.
Therefore, we complement the static approach by performing a series of first 
principles molecular dynamics (FPMD) simulations of the native oxide growth 
in the presence of surface dopants.
Apart from the choice of the initial conditions, dynamical simulations are 
unbiased regarding the possible reaction paths, and reveal mechanistic details 
otherwise very difficult to access experimentally.

\section{Methods}

Total energy calculations are performed within the spin-polarised
density functional theory, as implemented in the CASTEP code
\cite{CASTEP}.
The gradient-corrected exchange correlation functional generated by 
Perdew, Burke and Ernzerhof \cite{PBE} is employed, along with 
ultrasoft pseudopotentials to represent electron-ion interactions.
Wave functions are expanded in plane waves up to a kinetic energy
cut-off of 400~eV and a Gaussian smearing function of width 0.01~eV 
is applied to the electronic occupancies.
With this choice of parameters, energy differences were found to converge
to within 0.01~eV in selected test cases.
The equilibrium bond length and cohesive energy of bulk Si were
calculated to be 2.36~\AA\ and 4.55~eV/atom respectively (the
corresponding experimental values are 2.35~\AA\ and 4.63~eV/atom~
\cite{si_exper}).

Molecular dynamical simulations~\cite{Payne} are performed as above,
with a decreased plane wave energy cut-off of 300~eV and a time step of
1.5~fs.
Most of the simulations are performed within the microcanonical ensemble.
Where  temperature constraints are required, a Nos\'{e}-Hoover thermostat
with a 15~fs relaxation time is employed.
Convergence criteria for determining minimum energy structures are
set to $10^{-4}$~eV/atom for energy changes and 0.2~eV/\AA\ and
0.01~\AA{} for ionic forces and displacements respectively~\cite{Convergence}.
Atomic charges are computed via a Mulliken population analysis by projecting
the ground state wave functions onto a basis consisting of s and p atomic 
orbitals~\cite{Mull}.

Calculations on the Si(100) surface are performed using a periodically 
repeated $2\!\times\!2$  surface slab, sampling the Brillouin zone with two 
{\bf k}-points  at  $(0,\pm0.25, 0.25)$~\cite{Monk}. 
The surface model consists of 64 Si atoms arranged in eight layers and separated, 
in the $x$ direction, by a vacuum layer equivalent to a further eight layers. 
The surface energy per atom of such a slab is defined as:
$\gamma = {(E_{s}-E_{b})}/{n_{s}}$,
where $E_{s}$ is the total ground state energy of the 64-atom surface model,
$E_{b}$ is the energy of a correspondent bulk model without any vacuum
layer in the $x$ direction, and $n_{s}$ is the total number of surface atoms, 
16 in this case. 

\section{Reconstruction of the (100) Surface}

Cleaving a Si crystal  along the (100) plane leaves each surface atom 
with two dangling bonds and a high charge density between the surface 
atoms (Fig.~\ref{fig:1}~(left)). 
To stabilise this structure, the surface atoms relax to form a series of 
alternating buckled dimers with a length of $2.35$~\AA{} and a buckling 
angle of $21^{\circ}$~\cite{Kruger}, known as a p($2\!\times\!2$) reconstruction 
(Fig.~\ref{fig:1}~(right)).
The calculated surface energy, $\gamma$, of the reconstructed surface
is $1.15$~eV/atom~\cite{Wang}, consistent with the energy required to
break one bond per atom in bulk Si ($1.14$~eV/atom).
A Mulliken population analysis reveals that the dimer reconstruction
is accompanied by a redistribution of charge from the down dimer atom
(+0.14~$e^-$) to the up atom (-0.07~$e^-$). 
This charge separation corresponds to a change of orbital hybridisation of the 
down and up dimer atoms from sp$^{3}$ to sp$^{2}$ and p$^{3}$ respectively,
and has been shown to influence the pathway of oxygen chemisorption onto the 
bare surface~\cite{Lucio-oxide}.

The redistribution of electronic charge in the dangling bonds due to
cleavage of the crystal in the (100) plane results in the presence
of surface stress.
For symmetry reasons, this is diagonal in the reference frame with axes 
parallel ($\parallel$) and perpendicular ($\perp$) to the dimer bonds.
The two surface stress components are defined by the relations:
\begin{eqnarray*}
g_{\parallel} = \frac{1}{2}a\sigma_{\parallel},  \ \ \ 
g_{\perp} = \frac{1}{2}a\sigma_{\perp},
\end{eqnarray*}
where $\sigma_{\parallel}$ and $\sigma_{\perp}$ are the components 
of the stress tensor in the surface plane and $a$ is the height 
of the periodically repeated supercell~\cite{Van}. 
In all calculations, the lattice parameter in the surface plane has been kept
fixed to the bulk equilibrium value, and the stress component in the direction 
perpendicular to the surface was equal to zero.

The calculated $g$ components of the bulk-terminated and 
p($2\!\times\!2$)-reconstructed Si(100) surface are summarised and compared to 
previous ab-initio simulations~\cite{p2by2_stress} in Table~\ref{tab:1}. 
As thoroughly discussed in ~\cite{Haiss_2001}, upon truncation of the bulk,
the lone electrons tend to pull the surface atoms together to form
dimers in the $\parallel$ direction, resulting in high surface stress.
This is largely reduced by formation of buckled dimers, but a residual tension
remains, which is consistent with longer Si--Si bonds between atoms of the second 
and third layers beneath the dimer rows (Fig.~\ref{fig:1}). 
The bond angles at the down atom of each surface dimer are $120^\circ$, 
$119^\circ$ and $118^\circ$, close to ideal sp$^2$ bonding. 
However, the bond angle between the up dimer atom and its two neighbours
in the second layer is $99.5^\circ$, considerably larger than the  $90^\circ$
required for ideal p$^3$ bonding.
This leads to an increased tensile stress component $g_{\perp}$ in the 
p($2\!\times\!2$)-reconstructed surface with respect to the bulk truncated 
surface. 
Because of the positive stress anisotropy $g_{\parallel}-g_{\perp}$, alignment 
of the dimers in the direction of an externally applied compressive stress is expected
to be energetically favourable.
Indeed,  experiments show that the number of domains compressed along 
the dimer bond increases at the expense of domains compressed perpendicularly
to the bond~\cite{Men}.

\section{Oxidation of the (100) Surface}

We have seen in the previous section how structural features of the bare
surface reconstruction are linked to redistribution of charge and surface
stress, which we will show later to be important factors in controlling the 
segregation of impurities.
Under normal atmospheric conditions, however, the Si surface is
passivated by a thin oxide layer.
In ref.~\cite{Lucio-oxide}, the oxidation of the Si(100) surface in a
dry environment has been investigated by means of FPMD simulations. 
In good agreement with medium-energy ion scattering experiments~\cite{oxide_depth},
it was found that oxygen spontaneously adsorbs onto the bare surface up to a coverage of 
1.5 monolayers (ML).
At higher coverages, further oxide growth becomes limited by diffusion of O$_2$ 
molecules to the reactive Si/SiO$_x$ interface.
The charged Si species in the final structure obtained in~\cite{Lucio-oxide} after 
adsorption of  six O$_2$ molecules range from Si$^{+}$ to Si$^{4+}$.
We have performed a Mulliken analysis of this structure and found that 
the charges on the Si atoms increase approximately linearly with the number 
of nearest-neighbour O atoms, which act as electron acceptors (Fig.~\ref{fig:8}).

It is interesting to follow the development of stress in the partially oxidised 
surface at increasing oxygen coverage.
We have fully relaxed all structures obtained after each FPMD simulation 
in~\cite{Lucio-oxide}, keeping the lower surface of the slab 
p($2\!\times\!2$)-reconstructed.
The surface stress is then calculated as for the bare surface, except that, 
upon reaction with oxygen molecules, the symmetry of the dimer bonds 
is broken and shear components appear in the surface stress tensor.
The total surface stress, $g$, may be defined as the basis invariant trace of the 
stress tensor, $\sigma$, computed in a supercell of height $a$:
\begin{equation*}
g = \frac{1}{2}a\:\mbox{Tr}(\sigma_{\alpha\beta}) =
\frac{1}{2}a (\sigma_{yy}+\sigma_{zz}).\\
\end{equation*}
In order to isolate the development of stress in the oxide layer at the top of 
our surface slab, the total surface stress of the bottom p($2\!\times\! 2$) reconstructed
surface (0.81~Nm$^{-1}$) is subtracted from all of the results.

The evolution of the total surface stress as successive O$_2$ molecules 
react with the surface is summarised in Fig.~\ref{fig:9}.
The tensile stress initially present in the bare surface diminishes after
reaction of the first molecule.
In the final structure obtained, one O atom is adsorbed into a dimer bond, relieving 
the tension parallel to the bond direction (Fig.~\ref{fig:9}, red arrow).
The second O atom binds between  one up dimer atom and the neighbouring 
dimer  (Fig.~\ref{fig:9}, blue arrow), breaking a bond to the second 
layer (black arrow) and relaxing the tension in the direction perpendicular 
to the dimer.
Similar events lead to a further stress decrease after adsorption of a second
and a third oxygen molecule.
However, at a coverage of 1.25~ML, a backbone of alternating Si and O atoms 
develops in the surface oxide structure and extends over the neighbouring simulation
cells due to the periodic boundary conditions employed. 
Formation of this continuous oxide layer, given the constraints exerted by the
underlying crystalline bulk, results in a sudden appearance of a large tensile 
surface stress.

The obtained stress values are in reasonable agreement with the experimentally measured 
tensile stress of +0.26~Nm$^{-1}$ per monolayer of oxygen~\cite{Sander}. 
An absolute comparison is difficult, since in~\cite{Sander} the
authors assume a linear relationship between  oxygen coverage and surface 
stress, which is not observed here.
Moreover, the periodic boundary conditions which we apply to a small 
surface cell may exert a constraint on the extended network of Si--O bonds 
and result in an overestimation of the calculated stress.
However, this is not expected to affect either the observed stress evolution 
or the local environment of impurity atoms substituted into the oxide layer, 
which will be investigated in section~6.

The average Si--O bond in the fully oxidised structure is
$3\%$ longer than in the absence of surface stress, at a coverage of 0.75~ML.
This has been shown to drive dissociative water adsorption on
the native oxide layer~\cite{Lucio-water}.
Namely, stretching of the Si--O bond leads to a decrease in the Si d orbital
occupancies, allowing donation of water lone pairs onto electrophilic Si sites. 
Dissociative water adsorption promotes breaking of Si--O bonds in the oxide layer,
exposes reactive sites and thus facilitates further 
oxidation reactions~\cite{Lucio-water}.

\section{Impurity Segregation at Low Doping Concentration}

Before investigating the energetics of dopant substitution close to 
the Si(100) surface, single neutral P and B impurity atoms were substituted 
into a 64 atom bulk supercell. 
The close match in atomic radii between P and Si caused a
variation in bond length of less than 0.01~\AA{} and a negligible
decrease in energy on relaxation. 
Inspection of the density of states (DOS) of the doped system around the 
Fermi energy shows the presence of donor levels near the conduction band,
as expected (Fig.~\ref{fig:6}~(right)).
However, while the excess electron strongly localises close to the P atom,
a depletion of electron density in the bonding regions is visible from an
analysis of the charge density difference with respect to bulk Si
(Fig.~\ref{fig:6}~(left)). 
Indeed, a Mulliken population analysis reveals that the P atom actually 
receives some electronic charge from the surrounding atoms.

In contrast, B dopants are observed to add electron acceptor states close 
to the valence band and accept substantial electronic charge from the neighbouring
Si atoms, as shown in Fig.~\ref{fig:7}.
B has a much smaller atomic radius than both P and Si, so the dopant's 
four nearest-neighbours relax inwards by approximately 0.25~\AA\ 
upon structural relaxation. 
The displacement of the dopant itself is negligible. 
Previous theoretical work \cite{Nichols} also found little movement of 
the B atom, while its nearest-neighbours relaxed inwards by $0.2$~\AA. 

\subsection{P in the Bare Surface}

Starting with P dopants, we turn now to investigate the 
substitution of impurities close to the bare Si(100) surface.
As in the bulk, P substitution into the surface layers is not accompanied 
by significant atomic or charge redistribution. 
The close match in atomic radius between Si and P means that effects
due to stress are relatively unimportant, while the extra electron is 
mostly localised on the substitutional site.
The energy changes associated with the presence of a P dopant in the
first four surface layers with respect to the energy value upon 
substitution in the 64-atom bulk cell are shown in Fig.~\ref{fig:5}.
As far as the layers immediately below the surface are concerned, 
the small energy changes computed reveal no net driving force for 
P segregation from the bulk to subsurface sites. 
The small differences may be explained by considering the residual strain 
in the p($2 \times 2$) reconstruction.
In particular, the relative energetic stabilities of sites in the third and 
fourth layers appear to be correlated with the average bond length before 
substitution. 
Namely, in both layers the smaller P atom is unstable between the dimer rows, 
where average bond lengths are longer than the bulk Si bond length,
(position B  in Fig.~\ref{fig:1}), while the energy is 
reduced when the impurity is placed  beneath the rows of dimers, 
where average bond lengths are compressed relative to the bulk, 
(position A).

A significant  enhancement in stability over the bulk is obtained 
upon substitution of P for the upper atom of the surface dimer.
Indeed, the p$^3$ bonding features of this site mirror the chemical 
environment of P in molecules such as  PH$_3$ and PCl$_3$, in 
which the bond angles are $94^\circ$ and $100^\circ$ respectively. 
The stability of this threefold-coordinated site is supported by 
STM experiments where the presence of P--Si dimers has been revealed
after dissociation of phosphines and incorporation of P atoms in 
the (100) surface at 800~K~\cite{STMphosphine}.
Fig.~\ref{fig:4}~(centre) shows the environment and Mulliken charges around
the P atom on the most stable substitutional site, after relaxation. 
As in the bulk, the population analysis reveals that the dopant receives 
electrons from its neighbours and, hence, the electron depletion on the 
down dimer atom is enhanced over the bare surface. 

\subsection{B in the Bare Surface}

As in the case of P, two features of the B atom suggest that the 
minimum energy configuration will find it at, or near, the Si surface. 
Its small size causes a build up of stress in the Si lattice, which is reduced
after segregation to surface sites, and its electron acceptor properties lead to
saturation of surface dangling bonds.
The  energy of B dopants at low concentration as a function of depth below 
the Si ($100$) surface is shown in Fig.~\ref{fig:5}. 
Consistent with the evidence given by Ramamoorthy et al.~\cite{Ramamoo},
the minimum energy configuration corresponds to substitution into the 
second surface layer. 
As shown in Fig.~\ref{fig:4}~(right), this results in attraction of
electrons onto the B subsurface atom increasing the positive charge on
both neighbouring dimers.
When the B atom is in the first layer the energy is only slightly higher.
In this case, a large structural change is observed, in which the upper
atom in the buckled dimer moves towards the surface by nearly
$0.9$~\AA, flattening the dimer containing the impurity.

In slight contrast to the findings in~\cite{Ramamoo}, the energies of the
substitutional defects in layers three and four are already comparable with the bulk
value. 
As in the case of P, the small B atom is more stable on A sites 
relative to B sites.
Below the second layer, there is good correlation between the average
bond length before substitution and the relative energies of the B
impurity on that site. 
However, to explain the relatively low energies on surface sites, that 
are actually  in tension, chemical, rather than mechanical, effects 
must be considered. 
Namely, the small B atom favours sp$^2$ hybridisation and, indeed, the 
trend in energies reflects the extent to which the atoms around the impurity
can relax to this structure. 
B is most stable in the second layer, less stable in the first layer, where 
the bond angles are further from the ideal $120^{\circ}$, and least 
stable close to the bulk where the configuration is constrained to
 sp$^3$ bonding.

\section{Impurity Segregation in the Oxidised Si(100) Surface}

In this section we report the results of our study into the stability of
substitutional impurities at an oxidised Si(100) surface.
Our model of a thin native oxide layer on the (100) surface 
(Fig.~\ref{fig:9}) contains 1 Si$^{4+}$, 1 Si$^{3+}$, 6
Si$^{2+}$ and 5 Si$^{+}$ species. 
Three of the atoms initially present in the first and second surface
layers move below the surface and form no Si--O bonds. 
Stabilities of impurities in the oxidised surface were investigated by
substitution and structural relaxation of the dopants on all 16 of these sites.

\subsection{Phosphorus in the Native Oxide}

The dependence of the energy relative to the bulk substitution
on the number of O neighbours is shown in Fig.~\ref{fig:90}~(left)
for a single, neutral P atom.
The general trend reveals a decrease in stability as the number of O 
nearest-neighbours increases.
In other words, in the presence of a native oxide layer, P is expected 
to segregate preferentially to the Si side of the Si/SiO$_{x}$ interface.

Substitutions onto Si$^{4+}$ and Si$^{3+}$ sp$^3$ sites are accompanied
by very little relaxation. The surface tensile stress remains high and
the dopants are unstable. 
On Si$^{2+}$ sites there is a large spread in energies and this may be explained 
by observing the surface relaxation. 
At the most stable site, local structural rearrangements of the oxide
layer leave the substituted P atom in a favourable threefold-coordinated
configuration (black arrow in Fig.~\ref{fig:91}~(centre)).
Other rearrangements of two other Si$^{2+}$ sites leave the P atom 
sp$^3$-bonded but do reduce the tensile stress in the surface, thus 
slightly lowering the total energy. 
No such large-scale rearrangements are observed when P is placed on 
Si$^{+}$ sites, where the stability depends only on strain at the substitutional 
site. 
Bond lengths at the most stable site are actually compressed relative
to bulk Si, despite the high tensile stress in the oxide layer.

Despite the range of chemical and physical environments present in the
oxide layer, none of the sites discussed above are stable with respect
to bulk substitution. 
In contrast, the results show that segregation to any of the three 
non-oxidised sites at the Si/SiO$_x$ interface would be preferred. 
In the two most stable configurations, P replaces one Si atom of  
a very strained Si--Si bond (for example, red arrow in Fig.~\ref{fig:91}~(left)). 
Substitution of P further increases the separation, leaving 
the dopant favourably coordinated by three neighbours
close to an under-coordinated Si atom.
These results are consistent with a similar study of the
Si(100)/SiO$_2$ interface \cite{dabrow1}, in which it was found that
all structures containing P--O bonds were unstable by at least 0.5~eV in
sp$^3$ configurations or by 0.1~eV for threefold-coordinated P,
with respect to the bulk. 
In qualitative agreement with our finding,  Dabrowski et al. also report 
binding energies of 0.2~eV for sp$^3$ sites and of 1.1~eV for under-coordinated 
Si defect sites below the interface.

Our results imply that, at low doping concentrations, P should be
expelled from the native oxide layer and pile up on the Si side of the
interface. 
The most stable sites in our model leave an under-coordinated Si atom
after substitution.
It is likely that substitution at defect sites consisting of just one
under-coordinated Si atom would be characterised by a higher binding
energy.
Also, at higher doping concentrations, substitution at the remaining 
under-coordinated Si site to form a pair of dopants would be expected.

\subsection{Boron in the Native Oxide}

The effect of substituting a B atom onto sites in the native oxide layer is 
to increase the tensile surface stress by 0.31~Nm$^{-1}$, on average, and to 
deplete charge density on nearest-neighbour atoms.
Analysis of Mulliken charges reveals that B receives an approximately 
constant 0.8 electrons relative to Si, irrespective of the specific
substitution site.

The trend of calculated energy values with increasing numbers of O 
neighbours is similar to the case of P, as shown in Fig.~\ref{fig:90}~(right). 
In general, segregation to the bulk is preferred to the oxide layer,
especially to Si$^{4+}$ and Si$^{3+}$ sites. 
Despite the large size mismatch between B and Si, the large spread in
energies of the Si$^{2+}$ sites can be explained largely on the basis of
chemical (rather than mechanical) effects.
As in the bare surface, the low energy sites are those that allow B 
to form bond angles close to ideal sp$^2$ bonding. 
In the most stable site, however, the Mulliken charge of B is +1.25~$e^-$, 
which corresponds to leaving fewer than two electrons on the B atom. 
Indeed, the resulting structure is a linear O--B--O chain, in which B 
is sp hybridised (blue arrow, Fig.~\ref{fig:91}~(right)). 
The Si$^{+}$ sites are mostly very close in energy, with the exception 
of the compressed site discussed in the previous subsection.
The stability of this site is enhanced in the case of B since the final
bond lengths are much closer to the ideal B--Si length than on any
other site.

Given the stability of P at under-coordinated defect sites beneath
the native oxide layer, it is perhaps surprising that B is not as stable 
on such sites. 
This may be due to formation of a Si dangling bond upon B substitution
and breaking of the long Si--Si bond (dashed line in Fig.~\ref{fig:91}~(left)).
To investigate this issue, we replaced the under-coordinated Si atom
with a second B atom (Fig.~\ref{fig:82}). 
Despite this being an otherwise unstable Si$^{+}$ site, the energy of this 
configuration is 0.3~eV more stable than the bulk, indicating that B segregation 
to the native oxide interface may be enhanced by pairing of impurities.

Previous theoretical calculations of B impurities at a defect-free crystalline Si/SiO$_2$ 
interface~\cite{furu-theor} found a binding energy of 0.5~eV
on the Si side of the interface, while substitution into the oxide was
unstable by 2.2~eV.
This is in agreement with the energy we computed for the Si$^{4+}$ site. 
In the same paper, the authors find that introducing an O vacancy into the 
oxide allows the B atom to relax to an sp$^2$ configuration, but it is still
unstable with respect to the bulk by approximately 0.5~eV. 
Lowering the energy further by terminating the resulting Si dangling bond 
with H, they argue that B may segregate into the oxide layer by substituting onto
sites neighbouring O vacancies. 
Similarly, we find here that segregation of impurities is particularly 
favourable at under-coordinated sites on the Si side of the interface,
which exist at high concentration in the native oxide surface, and that
the interface may be stabilised by pairing of impurities.
Moreover, despite the general trend of avoiding forming B--O bonds,
we find that segregation to partially oxidised sites with compressed 
bond lengths may be favoured.

\section{Native Oxide Growth in the Presence of Surface Dopants}

In the previous section, we have performed static total energy calculations
with geometry optimisation to address the energetic stability of
substitutional P and B impurities in a pre-existing native oxide
layer model as obtained in FPMD simulations.
To substantiate the results of the static calculations we now perform
a series of FPMD simulations of the oxide formation on Si(100) 
surfaces initially doped with either P or B.
For each system, two impurity atoms have been placed in a 64-atom 
simulation cell, in minimum energy configurations predicted by the literature 
(Figs.~\ref{fig:1P}a and~\ref{fig:1B}a).
In particular, STM experiments~\cite{STMphosphine} reveal the presence 
of P--Si dimers at the bare surface, which we found to be more stable by approximately 
0.1~eV with respect to P--P dimers. 
In the case of B,  previous theoretical work predicts the segregation of pairs of B 
dopants  into the second layers, joined by two surface dimers, with negligible 
binding energy compared with isolated impurities~\cite{Ramamoo}. 

Initial oxygen chemisorption was investigated by placing two O$_2$ molecules
approximately 2.7~\AA{} above the surfaces, in the positions marked by
asterisks in Figs.~\ref{fig:1P}a and~\ref{fig:1B}a. 
In the case of the P-doped surface, only one of the two molecules was 
attracted towards the surface and dissociated after about 440~fs of simulation time 
 according to the hot-atom mechanism observed on bare Si surfaces~\cite{Lucio-oxide}
 (Fig.~\ref{fig:1P}b). 
The O atoms were inserted between a lower Si dimer atom and the second
layer, breaking a P--Si bond. 
The second O$_2$ molecule had not adsorbed after 750~fs, when the simulation 
was stopped. 
Chemisorption occurred more quickly on the B-doped surface. 
After 80~fs, both molecules were within 2~\AA{} of the surface and
later dissociated spontaneously (Fig.~\ref{fig:1B}b).

Fig.~\ref{fig:1P} shows further snapshots from the FPMD simulation 
of the oxidation of the P-doped surface  until the complete adsorption of
six O$_2$ molecules.
Characteristic features of the dynamics may be summarised as follows:
At low oxygen coverage, O$_{2}$ molecules bind preferentially to the
Si species, while they are repelled from the surface when placed in
starting positions immediately above the dopants.
P--O bonds form only occasionally and are, in part, observed to break 
during subsequent oxidation.
At intermediate coverages, molecules placed above charged Si species 
readily adsorb, but not necessarily dissociate immediately 
(as shown in Fig.~\ref{fig:1P}d).
Dissociation may take place later, forming both  bridging Si$^{2+}$ species 
and Si$^{4+}$ species at a coverage close to unity, as observed 
in~\cite{Lucio-oxide}.
In the final configuration obtained (Fig.~\ref{fig:1P}f), one P dopant
is bound to three Si neighbours, consistent with the thermodynamically
stable position found in the static total energy calculations.
The second P atom, however, is bound to two O atoms, most probably
trapped in a local energy minimum.
Diffusion processes inaccessible to the time-scale of our simulation may
help the atom escape the oxygen coordination. 
However, the possibility is not excluded that such metastable
structures may persist during oxidation at low temperature, such that
a small number of impurities may be found on the oxide side of the
Si/SiO$_{x}$ interface.

The development of a thin oxide film on the B-doped surface 
is shown in Fig.~\ref{fig:1B}.
The details of the simulations are generally similar to those found 
in the case of the bare surface or the P-doped system described
above.
Hot-atom dissociation of O$_{2}$ is observed to occur rapidly at low 
coverages upon binding to Si atoms neighbouring the B impurities.
At increasing coverage, Si$^{3+}$ species are observed to form while
the B dopants tend to adopt an sp$^2$ bonding configuration and to
remain segregated beneath the native oxide layer.
B--O bonds are observed to form temporarily, but never to last for
more than a few fs.
After reaching an oxygen coverage of approximately 1~ML, incoming oxygen
molecules are observed to chemisorb to the oxide structure and to remain 
undissociated for a few ps, after which the simulations were stopped.
In order to promote a phonon-mediated dissociation, the system is heated 
at a rate of 1~K per time step up to a temperature of 700~K.
Following this, the thermostat is switched off and the dynamics continued 
microcanonically.
Both molecules dissociated within 2~ps as a consequence of the
annealing.
The final structure obtained is shown in Fig.~\ref{fig:1B}f.

For both the P-doped and B-doped surfaces the simulations were
performed until an oxygen coverage of 1.5~ML was reached.
The final P-doped native oxide consists of 12 O atoms
bonded to 10 Si atoms, forming $30\%$ Si$^{4+}$, $10\%$ Si$^{3+}$,
$10\%$ Si$^{2+}$ and $50\%$ Si$^{+}$ species. 
In the B-doped structure, 12 O atoms were bonded to 
12 Si atoms, forming $8\%$ Si$^{4+}$, $25\%$ Si$^{3+}$, $17\%$
Si$^{2+}$ and $50\%$ Si$^{+}$ species. 
These compositions may be compared with that of the native oxide
layer obtained on the bare surface~\cite{Lucio-oxide} (12 O atoms 
bound to 13 Si atoms, forming $8\%$ Si$^{4+}$, $8\%$ Si$^{3+}$, 
$46\%$ Si$^{2+}$ and $38\%$ Si$^{+}$) and with the results of a 
synchrotron-radiation photoelectron spectroscopy study of a cleaned
Si(100) surface \cite{oxide_depth} (approximately 1:1 O:Si
stoichiometry, forming $10\%$ Si$^{4+}$, $14\%$ Si$^{3+}$, $25\%$
Si$^{2+}$ and $51\%$ Si$^{+}$).
During the simulations of the oxidation process, the average displacements 
of both P and B dopants perpendicular to the surfaces were typically 0.1~\AA{}.
The mean O atom depth is identical for the P- and B-doped systems. 
The P atoms lie 0.6~\AA{} and the B atoms 2.0~\AA{} below this mean
depth. 
This suggests a mechanism for oxidation whereby dopant atoms
remain at the Si/SiO$_x$ interface, while a native oxide layer
gradually builds up above them.

\section{Summary and Conclusions}

Given the importance of the electrostatic charge distribution of the natively
oxidised Si surface on the mediation of the interactions of Si-based devices
with their external environments, we have investigated the segregation
of P and B impurities during the oxidation of a Si(100) surface.
We have shown that segregation is controlled by a combination of 
chemical and mechanical effects.
In the case of the bare surface, these are illustrated by the influence 
of the surface reconstruction on the stability of P and B substitutional
defects.
With respect to substitution in the bulk, both impurities were found to 
be more stable by approximately 0.6~eV on chemically favourable surface 
sites, and to favour subsurface sites, in which the Si--Si bonds are 
compressed relative to the bulk.

Within the context of  impurity segregation, we have monitored
the changes in chemical and mechanical environments of the surface
during formation of a native oxide layer. 
It was found that surface tensile stress increased during this process, 
reaching 2.9~Nm$^{-1}$ for the surface saturated at a coverage of 1.5~ML. 
The native oxide layer contained a variety of oxidised Si species, whose 
Mulliken charges increase approximately linearly with the number of 
nearest-neighbour O atoms.
Substitution of Si by P and B impurities in the native oxide layer revealed 
that stable substitutional sites are located immediately below (for the case
of P) or at (for the case of B) the Si/SiO$_{x}$ interface.
FPMD simulations of oxide growth on doped bare Si(100) surfaces confirm
this general trend. 
Few P--O and B--O bonds were formed during the dynamics and segregation to 
threefold-coordinated sites was observed for both dopants. 
In the final structures, B is not bound to any oxygen atoms, 
while two P--O bonds are retained.

It is to be noted that neither a purely static nor a purely dynamic approach, 
by itself, is perfectly suited to simulate the real experimental situation.
On the one hand, the results of static total energy calculations may not represent 
entirely oxidation processes at low temperature, where the oxide structures may 
remain in metastable configurations.
On the other hand, FPMD simulations are intrinsically limited to local minimum
structures of the potential energy surface.
However, taken together,  the agreement between the two approaches used
here is remarkable given the very limited size of the studied system and the 
very short time addressable by the dynamical simulations.
The picture emerging from our results is indicative of an oxidation mechanism whereby 
the oxide layer grows above those dopants initially present on the bare surface, 
leaving them trapped at the Si/SiO$_x$ interface.
Dopants are therefore not expected to influence substantially the electrostatic and
Van der Waals interactions between natively oxidised Si surfaces and an external
atmospheric or liquid environment.
This is precious information for the construction of classical force fields to
be used in future simulations of realistic device surfaces on a larger scale. 

\subsection*{Acknowledgements}

Computational resources were provided by the HPCx supercomputing facilities
within the CPUK consortium, by the HPCf Cambridge, UK, by the Zentrum 
f\"ur Informationsdienste und Hochleistungsrechnen, Dresden, and by the
HLRS Stuttgart within the AQUOXSIM project and through the HPC-Europa 
project (RII3-CT-2003-506079, with the support of the  European Community - 
Research Infrastructure Action of the FP6).
LCC acknowledges support by the Alexander von Humboldt Stiftung and by
the Deutschen Forschungsgemeinschaft within the Emmy Noether Programme
(CI 144/2-1).
This work has been supported by the EPSRC, U.K.

\bibliographystyle{unsrt}
\bibliography{b_p_paper_03}

\begin{thebibliography}{10}
\bibitem{Convergence}
In selected test cases, we have lowered the convergence threshold on the forces
  to 0.05~eV/\AA{}, obtaining changes in energy differences smaller than
  0.05~eV. For instance, the differences in surface energy between a truncated
  and p($2\!\times\!2$) reconstructed Si surface are 1.153~eV/atom and
  1.149~eV/atom for thresholds of 0.2~eV/\AA{} and 0.05~eV/\AA{}, respectively.
  Moreover, the energy differences of a B impurity in the most stable
  substitution site with respect to the bulk are -0.75 and -0.78~eV for the two
  thresholds above.
\end{thebibliography}

\clearpage

\subsection*{Figure Captions}

\begin{figure}[h!]
\caption{The bulk-terminated (left) and  p($2\!\times\!2$)-reconstructed
(right) Si(100) surface.  The surface atoms pair to form dimers,
reducing the stress parallel to the bonds ($\parallel$). The
relatively large bond length between the second and third layers is
indicative of residual tension in this direction. The stress
perpendicular ($\perp$) to the dimer bonds originates from the large
angle between the up dimer atom and its second layer neighbours. In the
third and fourth layers, atoms lying on A sites are directly beneath
dimer rows, while the B sites are situated between the dimer rows. \label{fig:1} }
\end{figure}

\begin{figure}[h!]
\caption{Mulliken population analysis of central Si atoms with
increasing number of O nearest-neighbours. Charges are multiples of
the electronic charge. \label{fig:8}}
\end{figure}

\begin{figure}[h!]
\caption{Evolution of total surface stress, g (Nm$^{-1}$), during
growth of an oxide layer on the p($2\!\times\!2$)-reconstructed Si(100)
surface (top view) at increasing oxygen coverages, $\Theta$~(ML). O
atoms are shown in red and the model supercell is represented by the
dotted line. Dimer bonds between first layer atoms in the bare surface
are indicated by asterisks. Adsorption sites leading to decrease of stress
with respect to the bare surface are indicated by arrows (see text). 
\label{fig:9}}
\end{figure}

\begin{figure}[h!]
\caption{(left)
Charge-density difference associated with the substitution of a Si atom
by a P dopant in bulk Si. Positive isosurfaces, corresponding to
excess electrons on the P site and neighbouring atoms, are shown in
blue. Negative isosurfaces, corresponding to electron depletion in the
bonding regions, are shown in red. (right) Total density of states
of the system compared with bulk Si (dashed line). Donor levels are 
indicated by an arrow. (inset) Mulliken population analysis after 
structural relaxation, revealing that P receives electrons from surrounding 
atoms. Charges are multiples of the electronic charge.
\label{fig:6}}
\end{figure}

\begin{figure}[h!]
\caption{As in Fig.~\ref{fig:6}, for the case of B
substitution. (left) Charge-density difference with respect to bulk Si. 
(right) Total DOS compared with bulk Si (dashed line). Acceptor levels are indicated 
by an arrow. (inset) Mulliken charges (as multiples of the electronic charge) after 
structural relaxation. \label{fig:7}}
\end{figure}

\begin{figure}[h!]
\caption{Relative energy of a single, neutral, substitutional P atom
(left) and B atom (right) as a function of depth below the bare Si
($100$) surface. In the first layer, the energy is dependent on
whether the dopant is substituted into the upper or lower site of the
buckled dimer. In the third and fourth layers, the A sites are
situated directly beneath the dimer rows
while the B sites are situated between the dimer rows. 
Prior to substitution, the average Si--Si bond length at an A site
is shorter, and at a B site is longer, than the corresponding
bulk value.
The dotted line represents the energy of an impurity
in the bulk, deep below the Si surface. \label{fig:5}}
\end{figure}

\begin{figure}[h!]
\caption{(left) Mulliken charges (multiples of the electronic charge)
of selected atoms in the bare surface. Surface reconstruction is
accompanied by a charge transfer from the down dimer atom to the up
dimer. (centre) P segregates to the up dimer site, where the bond
angles are close to $90^\circ$, and further depletes electrons from
the down dimer. (right) B segregates to the second layer, where the bond
angles are close to the ideal $120^{\circ}$ for sp$^2$ bonding, and 
attracts electrons from all surrounding surface dimer atoms.\label{fig:4}}
\end{figure}

\begin{figure}[h!]
\caption{Relative energy of a single, neutral, substitutional P atom
(left) and B atom (right) as a function of number of nearest-neighbour
O in the oxide layer shown in Fig.~\ref{fig:9}. The general trend
reveals a tendency to avoid forming P--O and B--O bonds. The spread in
energies for a given number of nearest-neighbours may be explained by
combined strain and bonding effects. The dotted line represents the energy
of an impurity in the bulk, deep beneath the surface. \label{fig:90}}
\end{figure}

\begin{figure}[h!]
\caption{(left) 16 substitutional sites in the Si(100) native oxide
layer. Two Si$^{2+}$ sites are indicated by black and blue arrows. One
Si$^{+}$ site (green arrow) is under compression and is particularly
stable for both P and B dopants. The strained Si--Si bond (dashed line)
means that P and B are both stable on the site beneath the oxide layer
indicated by the red arrow. Both dopants move further away so that
they are threefold-coordinated and leave an under-coordinated Si
atom. (centre) Relaxed structure of the surface with P substituted
onto the Si$^{2+}$ site. The energy of substitution is reduced by
moving to a threefold-coordinated site. (right) Relaxed structure of
the surface with B substituted onto the most stable Si$^{2+}$ site,
forming a linear O--B--O chain. \label{fig:91}}
\end{figure}

\begin{figure}[h!]
\caption{Relaxed structure of the native oxide layer on Si(100) after
double B substitution on Si sites below the oxide layer (red arrow in
figure~\ref{fig:91}).\label{fig:82}}
\end{figure}

\begin{figure}[h!]
\caption{Side (parallel to the dimer rows) and top views of the
evolution of the P-doped Si(100) surface with increasing oxygen
coverage, $\Theta$. a) Starting configuration showing pairs of P atoms
(labelled 1 and 2) in upper dimer atom positions. Approximate
starting positions of O$_2$ molecules for the subsequent simulation
are indicated by asterisks. \label{fig:1P}}
\end{figure}

\begin{figure}[h!]
\caption{Side (parallel to the dimer rows) and top views of the
evolution of the B-doped Si(100) surface with increasing oxygen
coverage, $\Theta$. a) Starting configuration, showing pairs of B
atoms (labelled 1 and 2) in the second layer. Approximate starting
positions of O$_2$ molecules for the subsequent simulation are
indicated by asterisks. \label{fig:1B}}
\end{figure}

\clearpage

\subsection*{Tables}

\begin{table}[h!]
\caption{Components of the surface stress $g$ (Nm$^{-1}$) for the Si(100) surface
before and after reconstruction.~\label{tab:1}}
\begin{center}
\begin{tabular}{lccc}
&$g_{\parallel}$&$g_{\perp}$&$g_{\parallel}-g_{\perp}$\\[1mm]
\hline
\hline
\vphantom{\Large A}bulk-terminated&1.78&0.32&1.46\\[1mm]
p($2\!\times\!2$) (this work)&1.11&0.52&0.59\\[1mm]
p($2\!\times\!2$) \cite{p2by2_stress}&1.33&0.51&0.82\\[0.8mm]
\hline
\end{tabular}
\end{center}
\end{table}
\vfill\

\clearpage

\begin{center}
\includegraphics[width=5.15in]{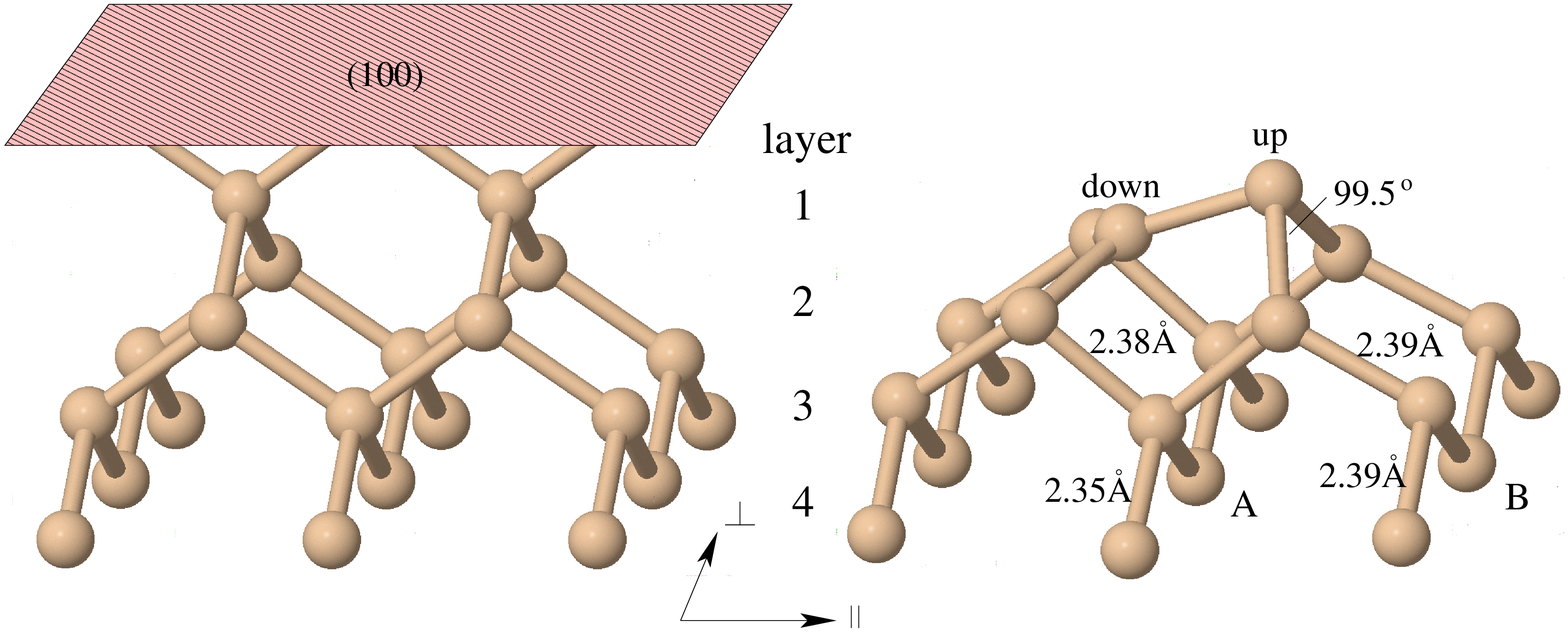}
\vfill
D. J. Cole, M. C. Payne, L. Colombi Ciacchi, Figure 1.
\end{center}

\clearpage

\begin{center}
\includegraphics[width=4in]{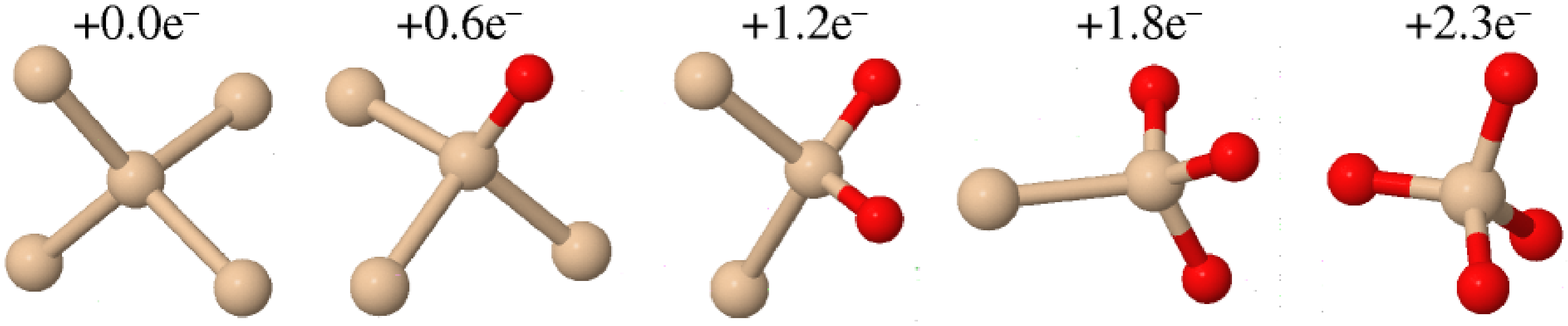}
\vfill
D. J. Cole, M. C. Payne, L. Colombi Ciacchi, Figure 2.
\end{center}

\clearpage

\begin{center}
\includegraphics[width=5.5in]{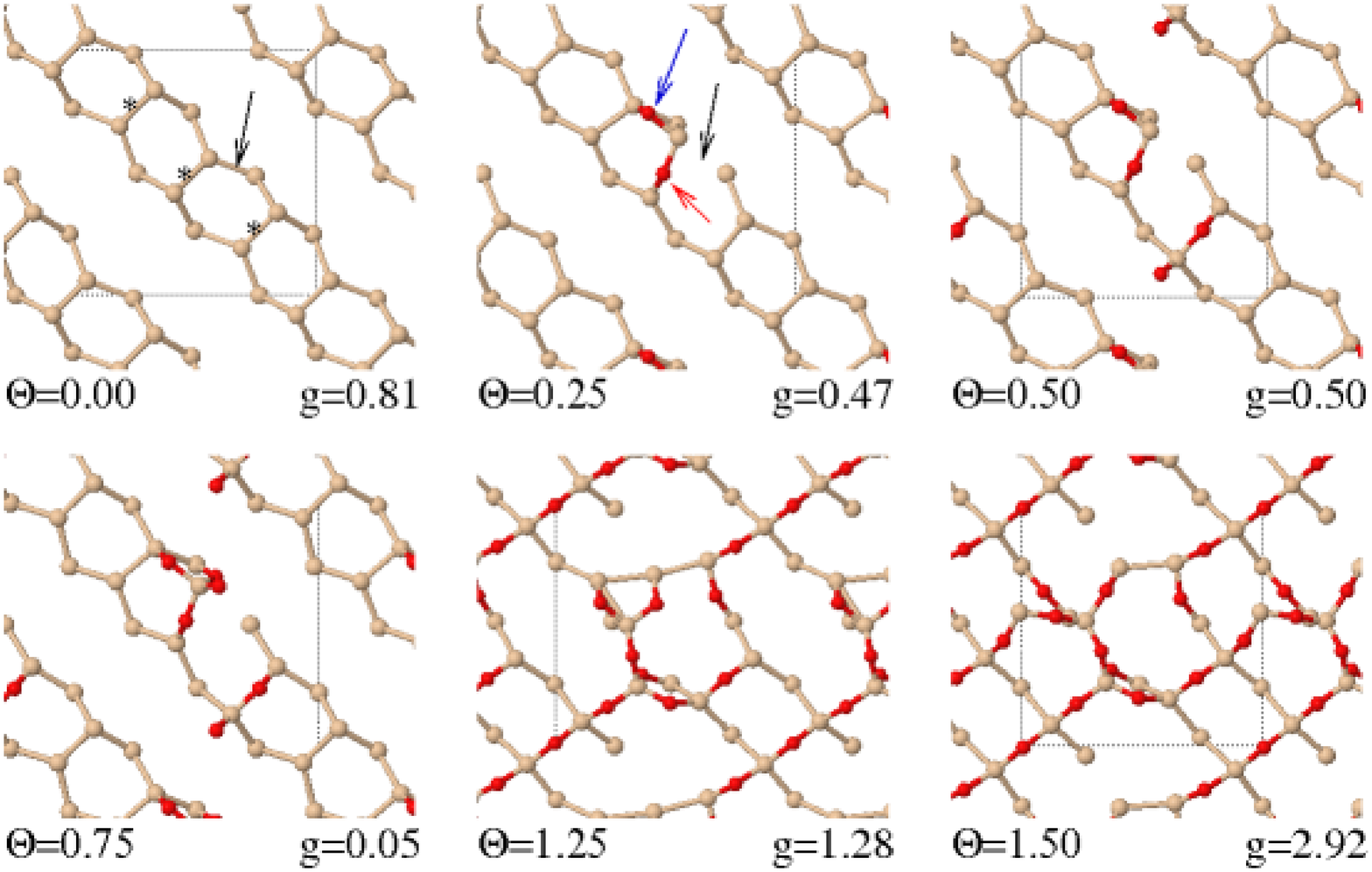}
\vfill
D. J. Cole, M. C. Payne, L. Colombi Ciacchi, Figure 3.
\end{center}

\clearpage

\begin{center}
\includegraphics[width=5in]{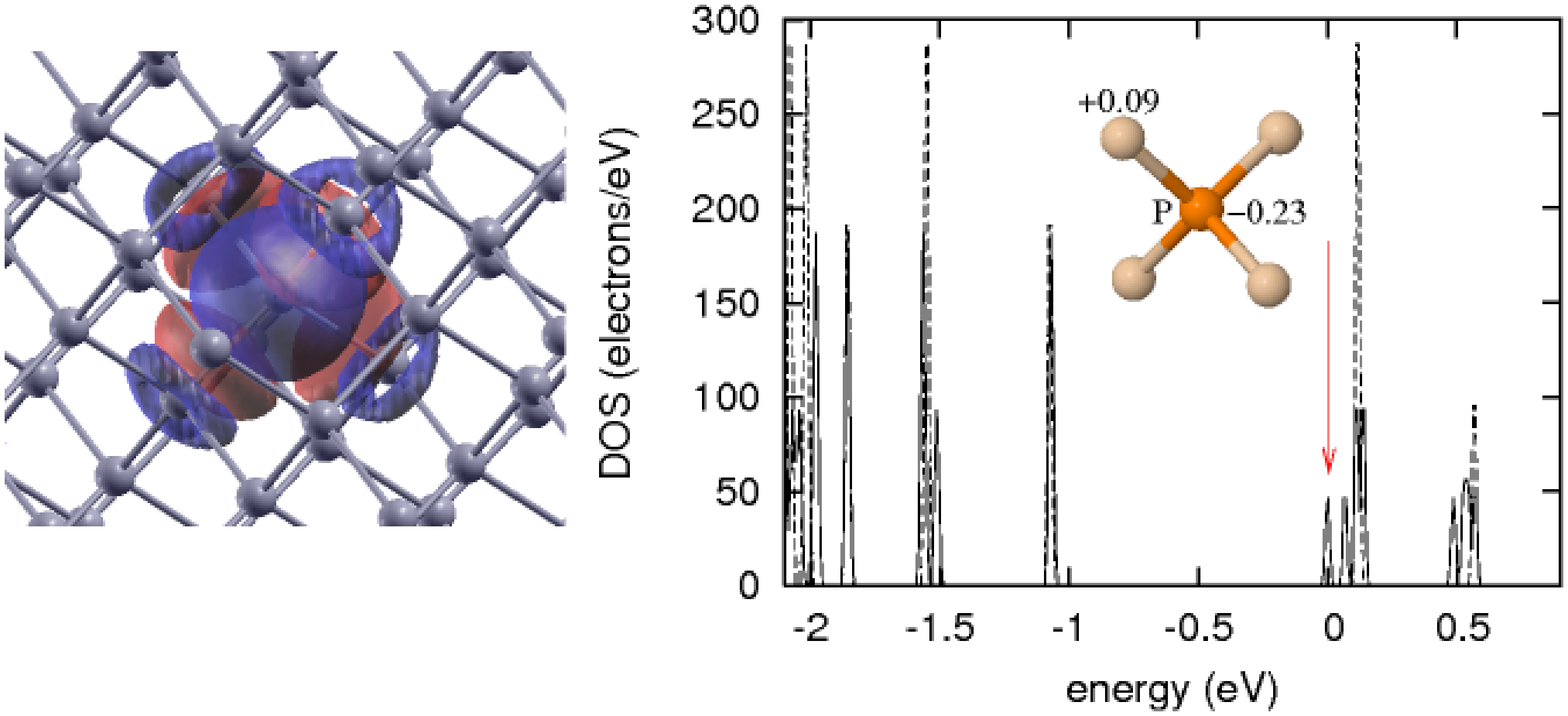}
\vfill
D. J. Cole, M. C. Payne, L. Colombi Ciacchi, Figure 4.
\end{center}

\clearpage

\begin{center}
\includegraphics[width=5in]{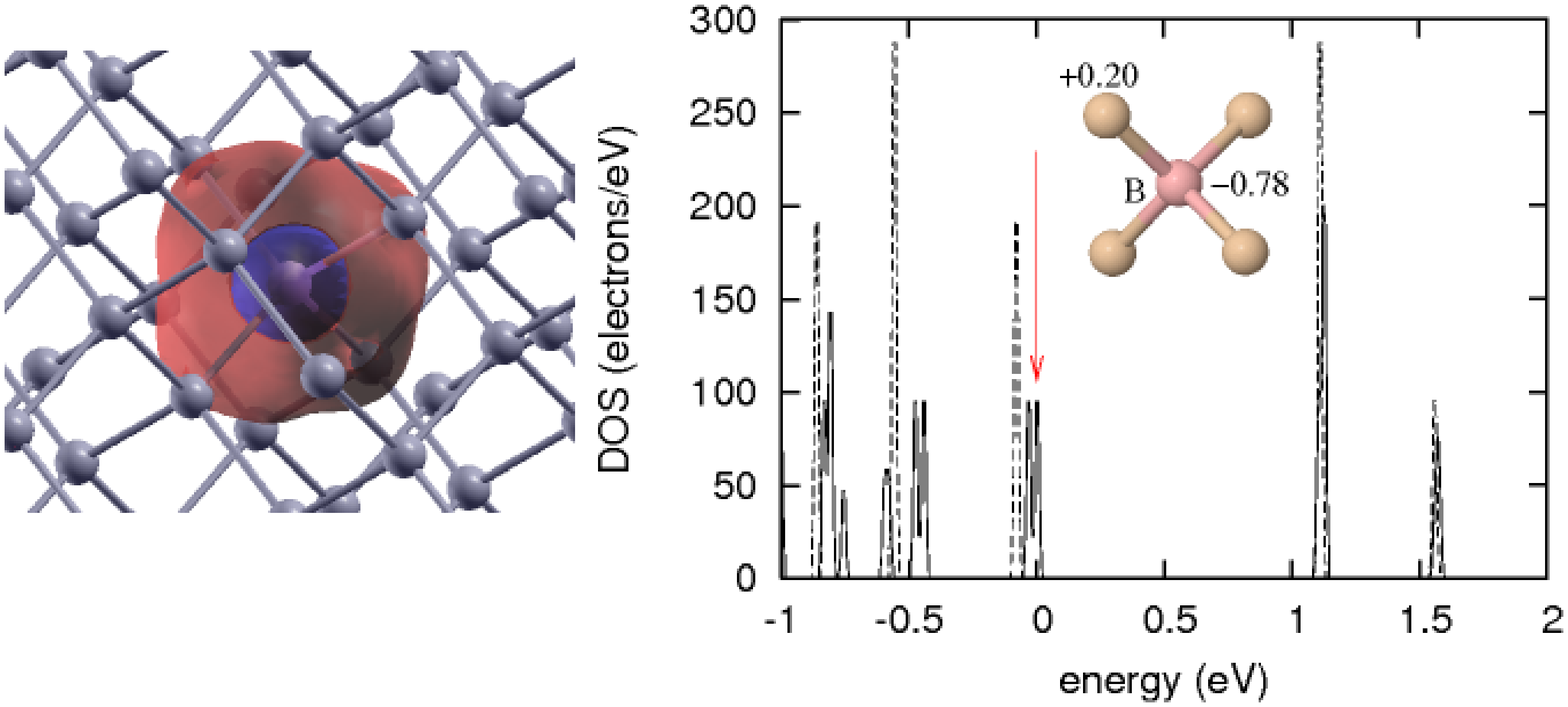}
\vfill
D. J. Cole, M. C. Payne, L. Colombi Ciacchi, Figure 5.
\end{center}

\clearpage

\begin{center}
\includegraphics[width=3in]{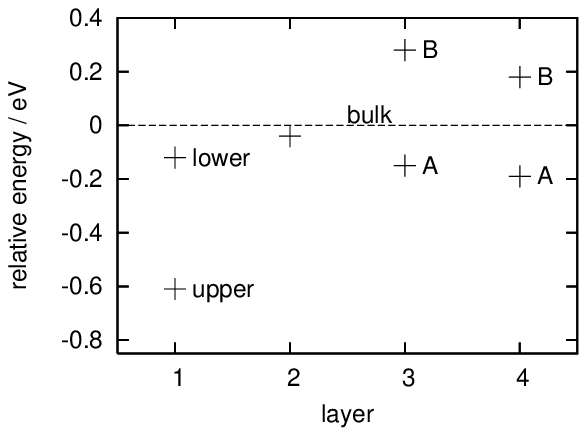}
\includegraphics[width=3in]{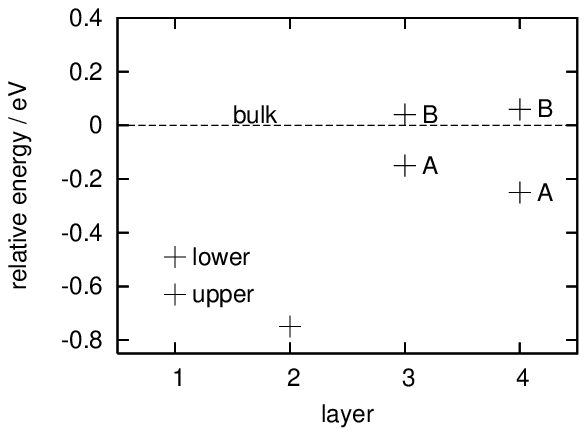}
\vfill
D. J. Cole, M. C. Payne, L. Colombi Ciacchi, Figure 6.
\end{center}

\clearpage

\begin{center}
\includegraphics[width=5in]{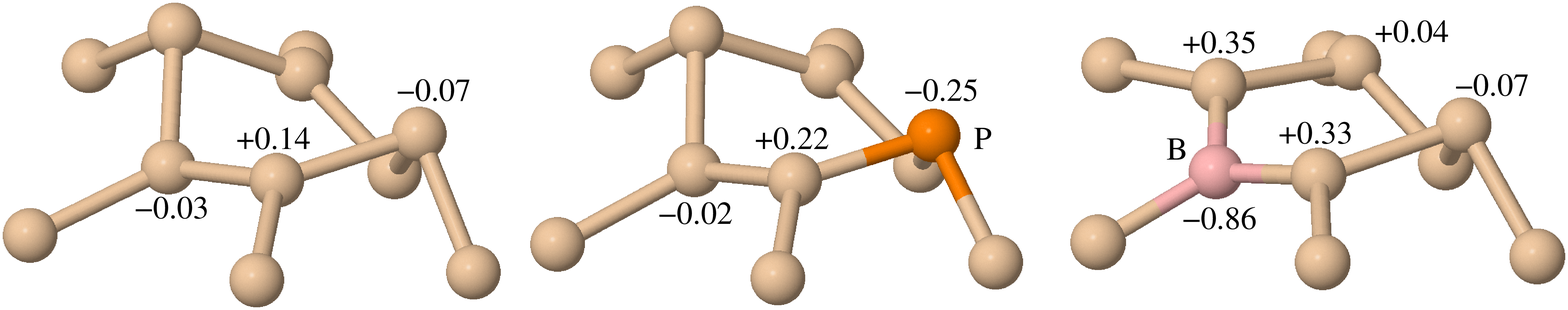}
\vfill
D. J. Cole, M. C. Payne, L. Colombi Ciacchi, Figure 7.
\end{center}

\clearpage

\begin{center}
\includegraphics[width=3in]{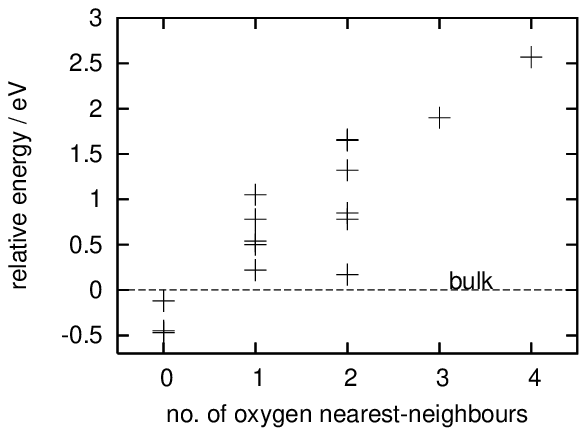}
\includegraphics[width=3in]{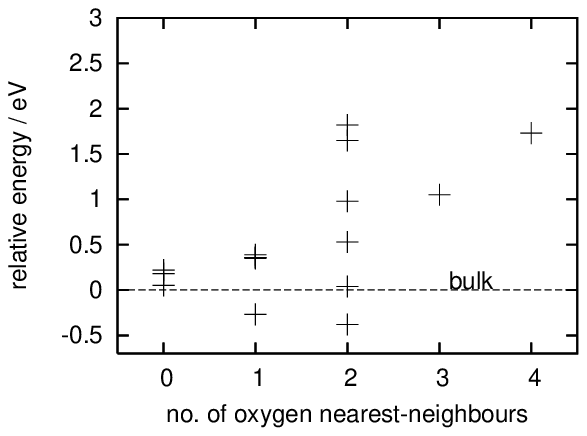}
\vfill
D. J. Cole, M. C. Payne, L. Colombi Ciacchi, Figure 8.
\end{center}

\clearpage
\begin{center}
\includegraphics[width=5.9in]{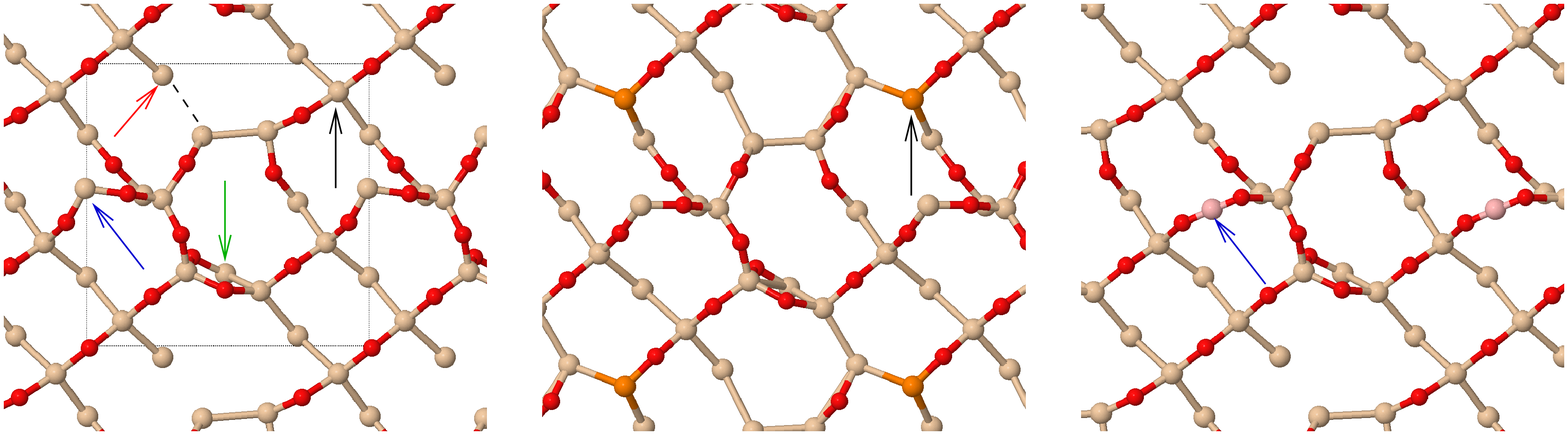}
\vfill
D. J. Cole, M. C. Payne, L. Colombi Ciacchi, Figure 9.
\end{center}

\clearpage

\begin{center}
\includegraphics[width=2in]{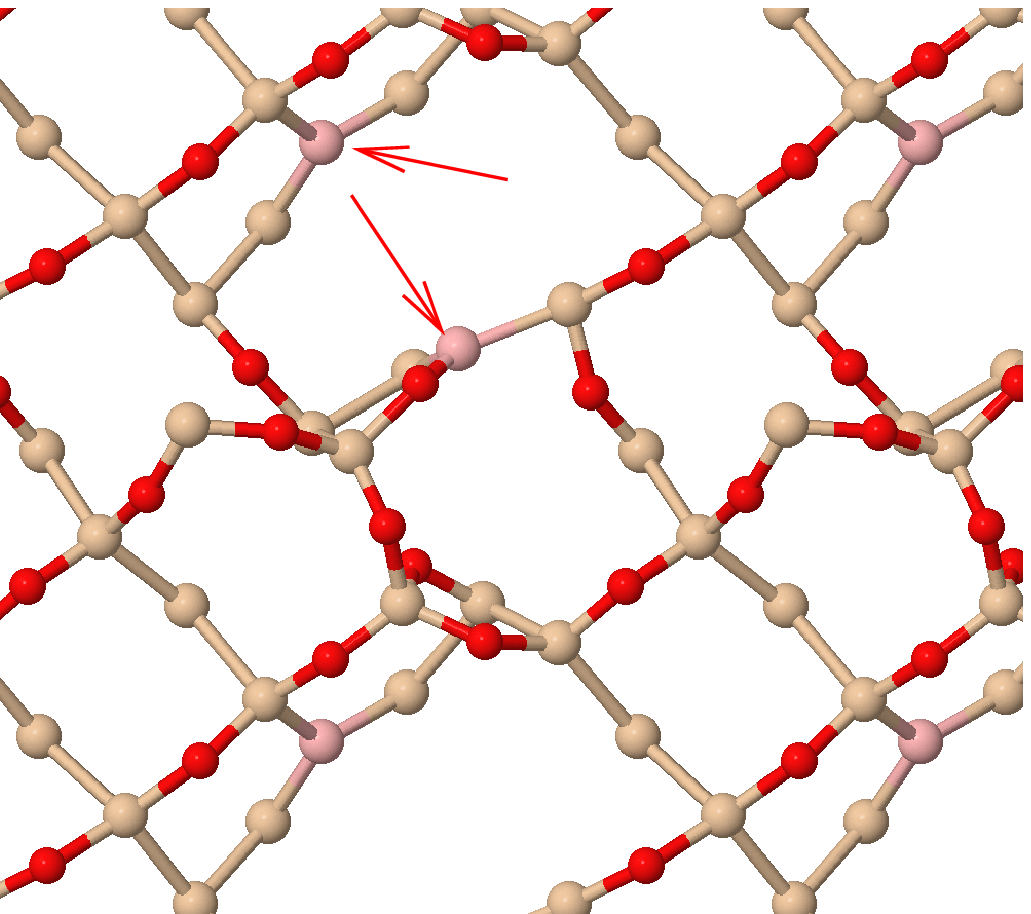}
\vfill
D. J. Cole, M. C. Payne, L. Colombi Ciacchi, Figure 10.
\end{center}

\clearpage

\begin{center}
\includegraphics[width=5.9in]{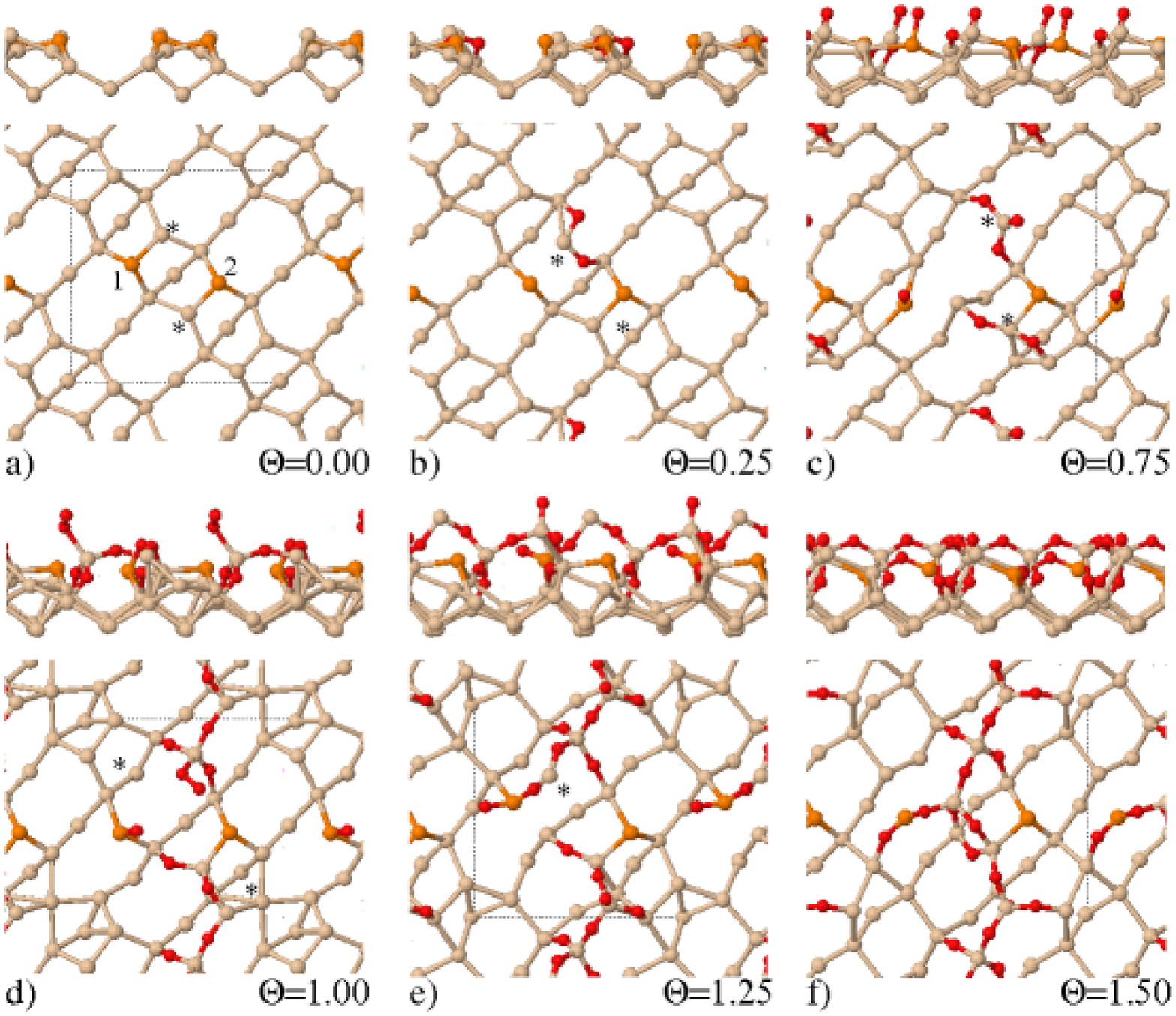}
\vfill
D. J. Cole, M. C. Payne, L. Colombi Ciacchi, Figure 11.
\end{center}

\clearpage

\begin{center}
\includegraphics[width=5.9in]{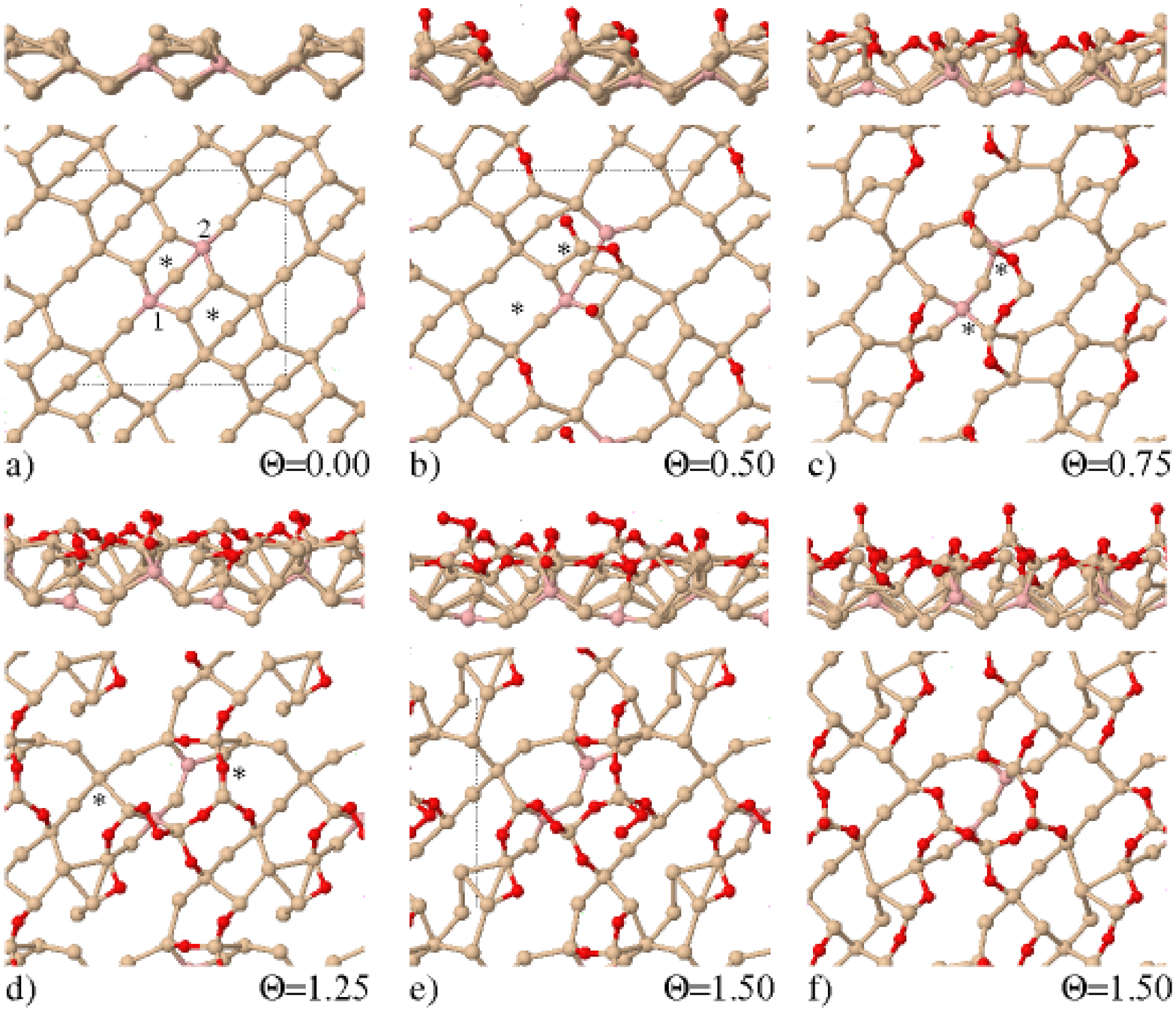}
\vfill
D. J. Cole, M. C. Payne, L. Colombi Ciacchi, Figure 12.
\end{center}

\end{document}